\documentclass[prd]{revtex4}

\usepackage{amsmath,amssymb}
\usepackage{graphicx}
\usepackage{pst-plot}
\usepackage{mathrsfs}
\usepackage{mathtools}
\usepackage{pst-func}
\usepackage{rotating}
\usepackage{epsfig}

\usepackage{chemarrow}

\usepackage[titletoc]{appendix}

\begin{document}

\title{Macroscopic Observability of Spinorial Sign Changes: A Reply to Gill}

\author{Joy Christian}

\email{jjc@alum.bu.edu}

\affiliation{Einstein Centre for Local-Realistic Physics, 15 Thackley End, Oxford OX2 6LB, United Kingdom}

\begin{abstract}
In a recent paper Richard Gill has criticized an experimental proposal published in a journal of physics which describes how to detect a macroscopic signature of spinorial sign changes. Here I point out that Gill's worries stem from his own elementary algebraic and conceptual mistakes, and present\break several event-by-event numerical simulations which bring out his mistakes by explicit computations.
\end{abstract}

%\pacs{04.20.Gz, 04.20.Cv}

\maketitle

\parskip 5pt

In a recent paper a mechanical experiment has been proposed to test possible macroscopic observability of spinorial sign changes under ${2\pi}$ rotations \cite{IJTP}. The proposed experiment is a variant of the local model for the spin-1/2 particles considered by Bell \cite{Bell}, which was later further developed by Peres providing pedagogical details \cite{Peres}\cite{Can}. Our experiment differs, however, from the one considered by Bell and Peres in one important respect. It involves measurements of the actual spin angular momenta of two fragments of an exploding bomb rather than their normalized spin values, ${\pm 1}$.

It is well known that angular momenta are best described, not by ordinary polar vectors, but by pseudo-vectors, or bivectors, that change sign upon reflection \cite{Can}\cite{Christian}. One only has to compare a spinning object, like a barber's pole, with its image in a mirror to appreciate this elementary fact.
The mirror image of a polar vector representing the spinning object is not the polar vector that represents the mirror image of the spinning object. In fact it is the negative of the polar vector that does the job. Therefore the spin angular momenta ${{\bf L}({\bf a},\,\lambda)}$ in the theoretical analysis of the experiment proposed in the paper \cite{IJTP} have been represented by bivectors using the powerful language of geometric algebra \cite{GA}. They can be expressed in terms of the bivector basis (which are {\it graded} basis) satisfying the sub-algebra
\begin{equation}
L_{\mu}(\lambda)\,L_{\nu}(\lambda) \,=\,-\,g_{\mu\nu}\,-\,\sum_{\rho}\,\epsilon_{\mu\nu\rho}\,L_{\rho}(\lambda)\,, \label{wh-o8899}
\end{equation}
where ${\lambda=\pm\,1}$ represents a choice of orientation of a unit 3-sphere \cite{IJTP}. This brings us to the first of several elementary mistakes made by Gill. In his paper \cite{Gillprint} he claims that, from the above equation, using ${L_{\mu}(\lambda=-1)=-\,L_{\mu}(\lambda=+1)}$, 
\begin{align}
&L_{\mu}(\lambda=+1)\,L_{\nu}(\lambda=+1)\,-\,L_{\mu}(\lambda=-1)\,L_{\nu}(\lambda=-1) \,=\,-\,\sum_{\rho}\,\epsilon_{\mu\nu\rho}\,L_{\rho}(\lambda=+1)\,+\,\sum_{\rho}\,\epsilon_{\mu\nu\rho}
\,L_{\rho}(\lambda=-1) \label{2}\\
&\text{implies}\;\;\;\;\;\;
0\,=\,-2\sum_{\rho}\,\epsilon_{\mu\nu\rho}\,L_{\rho}(\lambda=+1)\,=\,+2\sum_{\rho}\,\epsilon_{\mu\nu\rho}\,L_{\rho}(\lambda=-1),
\label{notwh-o8899} \\
&\text{which in turn implies}\;\;\;\;\;\;
L_{\rho}(\lambda=+1)\,=\,L_{\rho}(\lambda=-1) \,=\,0\,.\label{yeswh-o8899}
\end{align}
It is not difficult to see, however, that this claim is manifestly false. Gill's mistake here is to miss the summation over the index ${\rho}$. And since the basis elements ${L_{\rho}(\lambda=+1)}$ and ${L_{\rho}(\lambda=-1)}$ do not vanish in general, contrary to his claim neither the spin angular momentum ${{\bf L}({\bf a},\,\lambda)}$ about the direction ${\bf a}$ nor the spin angular momentum ${{\bf L}({\bf b},\,\lambda)}$ about the direction ${\bf b}$ vanish in general. Only the spin angular momentum ${{\bf L}({\bf a}\times{\bf b},\,\lambda)}$ about the mutually orthogonal direction ${{\bf a}\times{\bf b}}$ vanish, as in Eq.${\,}$(110) of the paper \cite{IJTP}. This is of course consistent with the physical fact that there is no third fragment of the bomb spinning about the direction ${{\bf a}\times{\bf b}}$ exclusive (as well as orthogonal) to both directions ${\bf a}$ and ${\bf b}$. To see that only the spin angular momentum ${{\bf L}({\bf a}\times{\bf b},\,\lambda)}$ about the direction ${{\bf a}\times{\bf b}}$ vanishes, all one has to do is to contract Eq.${\,}$(\ref{wh-o8899}) above with the vector components ${a_{\mu}}$ and ${b_{\nu}}$, on both sides, and then follow through the steps in Eqs.${\,}$(\ref{2}) and (\ref{notwh-o8899}). It is also crucial to appreciate that the spin angular momenta ${{\bf L}({\bf s},\,\lambda)}$ ({\it i.e.}, the bivectors) trace out an su(2) 2-sphere within the group manifold ${{\rm SU(2)}\sim S^3}$, not a round ${S^2}$ within ${{\rm I\!R^3}}$ as Gill has incorrectly assumed.  

Unfortunately the conceptual mistakes made by Gill are even more serious than the above elementary mathematical mistakes \cite{Gillprint}\cite{refute}. As already noted, what are supposed to be measured in the proposed experiment are the spin angular momenta ${{\bf L}({\bf a},\,\lambda)}$ and ${{\bf L}({\bf b},\,\lambda)}$ themselves, not their normalized spin values ${\pm 1}$ about some directions ${\bf a}$ and ${\bf b}$, where
\begin{align}
{\bf L}({\bf a},\,{\lambda})\,&\equiv\,A({\bf a},\,\lambda)\,\equiv\,\lambda({\bf e}_x\wedge{\bf e}_y\wedge{\bf e}_z)\cdot{\bf a}\;=\,\pm 1 \;\,\text{spin about}\;\,
{\bf a} \\
\text{and}\;\;\;{\bf L}({\bf b},\,{\lambda})\,&\equiv\,B({\bf b},\,\lambda)\,\equiv\,\lambda({\bf e}_x\wedge{\bf e}_y\wedge{\bf e}_z)\cdot{\bf b}\,=\,\pm 1 \;\,\text{spin about}\;\, {\bf b}\,.
\end{align}
This is in sharp contrast to what are measured as dynamical variables in the model considered by Bell and Peres \cite{Can}.

In practice the above dynamical variables are supposed to be measured in the proposed experiment by directly observing the polar vectors ${{\bf s}^k}$ {\it dual} to the random bivectors ${{\bf L}({\bf s},\,\lambda^k)}$. It is only after all the runs of the experiment are completed and the vectors ${{\bf s}^k}$ are fully recorded, the traditional dynamical variables ${{sign}\,(\,+\,{\bf s}^k\cdot{\bf a})}$ and ${{sign}\,(\,-\,{\bf s}^k\cdot{\bf b})}$ supposed by Bell are to be calculated, by an algorithm extraneous to the actual experiment. These calculations may be done, for example, years after the experiment has been completed. Thus they are not supposed to be an integral part of the physical experiment itself, and this fact must be accounted for in the ensuing statistical analysis of data.

Since Gill has failed to appreciate the difference between the actual dynamical variables in the proposed experiment, namely the bivectors ${{\bf L}({\bf a},\,\lambda^k)}$ and ${{\bf L}({\bf b},\,\lambda^k)}$, and the subsequently calculated traditional variables ${{sign}\,(\,+\,{\bf s}^k\cdot{\bf a})}$ and ${{sign}\,(\,-\,{\bf s}^k\cdot{\bf b})}$, he has ended up misapplying his calculation, which leads him to a result not relevant to the problem at hand. In order to correctly calculate the correlation between the variables ${{sign}\,(\,+\,{\bf s}^k\cdot{\bf a})}$ and ${{sign}\,(\,-\,{\bf s}^k\cdot{\bf b})}$ one must first specify their relation to the actual dynamical variables ${{\bf L}({\bf a},\,\lambda^k)}$ and ${{\bf L}({\bf b},\,\lambda^k)}$ to be observed in the experiment:
\begin{align}
S^3\ni\pm1\,=\,{sign}\,(\,+\,{\bf s}^k\cdot{\bf a})\,\equiv\,
{\mathscr A}({\bf a},\,{\lambda^k})\,&=\,\lim_{{\bf s}\,\rightarrow\,{\bf a}}\left\{\,-\,{\bf D}({\bf a})\,{\bf L}({\bf s},\,\lambda^k)\right\}
=\,                  
\begin{cases}
+\,1\;\;\;\;\;{\rm if} &\lambda^k\,=\,+\,1 \\
-\,1\;\;\;\;\;{\rm if} &\lambda^k\,=\,-\,1
\end{cases} \label{88-oi}
\end{align}
and
\begin{align}
S^3\ni\mp1\,=\,{sign}\,(\,-\,{\bf s}^k\cdot{\bf b})\,\equiv\,
{\mathscr B}({\bf b},\,{\lambda^k})\,&=\,\lim_{{\bf s}\,\rightarrow\,{\bf b}}\left\{\,+\,{\bf D}({\bf b})\,{\bf L}({\bf s},\,\lambda^k)\right\}
=\,                     
\begin{cases}
-\,1\;\;\;\;\;{\rm if} &\lambda^k\,=\,+\,1 \\
+\,1\;\;\;\;\;{\rm if} &\lambda^k\,=\,-\,1\,,
\end{cases} \label{99-oi}
\end{align}
where the orientation ${\lambda}$ of ${S^3}$ is assumed to be a random variable with 50/50 chance of being ${+1}$ or ${-\,1}$ at the moment of the bomb-explosion, making the spinning bivector 
${{\bf L}({\bf a},\,\lambda)}$ a random variable {\it relative} to the detector bivector ${{\bf D}({\bf a})}$:
\begin{equation}
{\bf L}({\bf a},\,\lambda)
\,\equiv\,\{\,a_{\mu}\;L_{\mu}(\lambda)\,\}\,=\,\lambda\,\{\,a_{\nu}\;D_{\nu}\,\}\,\equiv\,\lambda\,{\bf D}({\bf a}). \label{OJS}
\end{equation}

From the above discussion it should be clear that the raw scores ${{sign}\,(\,+\,{\bf s}^k\cdot{\bf a})}$ and ${{sign}\,(\,-\,{\bf s}^k\cdot{\bf b})}$ would be generated in the experiment with {\it different} bivectorial scales of dispersion, or {\it different} standard deviations, as explained between Eqs.${\,}$(102) and (109) in Ref.${\,}$\cite{IJTP}. Therefore the calculation of the correlation between these raw scores, as well as the derivation of the Tsirel'son's bound on the strength of possible correlations, must be carried out with some care. In fact there are at least four physical considerations one must be mindful of before proceeding further: (1) the spin angular momenta are represented, not by polar vectors, but by bivectors that change sign upon reflection; (2) scalars and bivectors, despite being elements of different grades, are treated on equal footing in geometric algebra; (3) what would be actually observed in the proposed experiment are not the raw scores ${{\mathscr A}({\bf a},\,{\lambda}^k)\equiv{sign}\,(\,+\,{\bf s}^k\cdot{\bf a})}$ and ${{\mathscr B}({\bf b},\,{\lambda}^k)\equiv{sign}\,(\,-\,{\bf s}^k\cdot{\bf b})}$ but the standard scores ${A({\bf a},\,\lambda^k)\equiv{\bf L}({\bf a},\,\lambda^k)}$ and ${B({\bf b},\,\lambda^k)\equiv{\bf L}({\bf b},\,\lambda^k)}$; and (4) the correct association between the raw scores ${{\mathscr A}({\bf a},\,{\lambda}^k)}$ and ${{\mathscr B}({\bf b},\,{\lambda}^k)}$ can be inferred only by calculating the covariance between the corresponding standard scores ${A({\bf a},\,\lambda^k)}$ and ${B({\bf b},\,\lambda^k)}$. With these in mind, let us consider four reference vectors ${\bf a}$,\break ${\bf a'}$, ${\bf b}$, and ${\bf b'}$. Then the bound on the corresponding CHSH string of expectation values \cite{Christian}, namely, on the coefficient
\begin{equation}
{\cal E}({\bf a},\,{\bf b})\,+\,{\cal E}({\bf a},\,{\bf b'})\,+\,
{\cal E}({\bf a'},\,{\bf b})\,-\,{\cal E}({\bf a'},\,{\bf b'})\,, \label{B1-11}
\end{equation}
can be derived using the four joint expectation values of the raw scores --- such as
${{\mathscr A}({\bf a},\,{\lambda})}$ and ${{\mathscr B}({\bf b},\,{\lambda})}$ --- defined as
\begin{equation}
{\cal E}({\bf a},\,{\bf b})\,=\lim_{\,n\,\gg\,1}\left[\frac{1}{n}\sum_{k\,=\,1}^{n}\,
{\mathscr A}({\bf a},\,{\lambda}^k)\;{\mathscr B}({\bf b},\,{\lambda}^k)\right]\,\equiv\,\Bigl\langle\,{\mathscr A}_{\bf a}({\lambda})\,{\mathscr B}_{\bf b}({\lambda})\,\Bigr\rangle\,.\label{exppeu}
\end{equation}
This allows us to express the above CHSH string of expectation values simply as a string of four averages as follows:   
\begin{equation}
\Bigl\langle\,{\mathscr A}_{\bf a}({\lambda})\,{\mathscr B}_{\bf b}({\lambda})\,\Bigr\rangle\,+\, \Bigl\langle\,{\mathscr A}_{\bf a}({\lambda})\,{\mathscr B}_{\bf b'}({\lambda})\,\Bigr\rangle\,+\, \Bigl\langle\,{\mathscr A}_{\bf a'}({\lambda})\,{\mathscr B}_{\bf b}({\lambda})\,\Bigr\rangle\,-\, \Bigl\langle\,{\mathscr A}_{\bf a'}({\lambda})\,{\mathscr B}_{\bf b'}({\lambda})\,\Bigr\rangle\,.\label{four}
\end{equation}
It is in the next step that Gill makes his gravest mistake. He surreptitiously replaces the above string of four separate averages of numbers that are generated with {\it different} bivectorial scales of dispersion with the following single average:
\begin{equation}
\Bigl\langle\,{\mathscr A}_{\bf a}({\lambda})\,{\mathscr B}_{\bf b}({\lambda})\,+\,
{\mathscr A}_{\bf a}({\lambda})\,{\mathscr B}_{\bf b'}({\lambda})\,+\, {\mathscr A}_{\bf a'}({\lambda})\,{\mathscr B}_{\bf b}({\lambda})\,-\,
{\mathscr A}_{\bf a'}({\lambda})\,{\mathscr B}_{\bf b'}({\lambda})\,\Bigr\rangle\,.\label{one}
\end{equation}
As innocuous as this step may seem, it is in fact an illegitimate mathematical step within the context of the experiment proposed in Ref.${\,}$\cite{IJTP}. But this illegitimate maneuver does allow Gill to reduce the above average at once to the average 
\begin{equation}
\Bigl\langle\,{\mathscr A}_{\bf a}({\lambda})\,\big\{{\,\mathscr B}_{\bf b}({\lambda})+{\mathscr B}_{\bf b'}({\lambda})\,\big\}\,+\,{\mathscr A}_{\bf a'}({\lambda})\,\big\{\,{\mathscr B}_{\bf b}({\lambda})-{\mathscr B}_{\bf b'}({\lambda})\,\big\}\,\Bigr\rangle\,. 
\end{equation}
And since ${{\mathscr B}_{\bf b}({\lambda})=\pm1}$, if ${|{\mathscr B}_{\bf b}({\lambda})+{\mathscr B}_{\bf b'}({\lambda})|=2}$, then ${|{\mathscr B}_{\bf b}({\lambda})-{\mathscr B}_{\bf b'}({\lambda})|=0}$, and vice versa. Consequently, using ${{\mathscr A}_{\bf a}({\lambda})=\pm1}$, it is easy to conclude that the absolute value of the above average cannot exceed 2, as Gill has concluded. 

As compelling as this conclusion by Gill may seem at first sight, it is entirely false. It is based on his illegitimate and careless maneuver of replacing the string of four separate averages of random variables (\ref{four}) with a single average (\ref{one}). Such a move is justified on mathematical grounds {\it only} if all four random variables ${{\mathscr A}_{\bf a}({\lambda})}$, ${{\mathscr B}_{\bf b}({\lambda})}$, ${{\mathscr A}_{\bf a'}({\lambda})}$, and ${{\mathscr B}_{\bf b'}({\lambda})}$ are on equal statistical and geometrical footings. But as we noted earlier, each one of these variables is generated with a {\it different} bivectorial scale of dispersion, or a {\it different} standard deviation, as explained between equations (102) and (109) in Ref.${\,}$\cite{IJTP}. Therefore the {\it theoretical} prediction of the correlation (\ref{exppeu}) between these raw scores, as well as the {\it theoretical} derivation of the Tsirel'son's bound on the strength of possible correlations, {\it must} proceed as follows.

Since the correct association between the raw scores ${{\mathscr A}({\bf a},\,{\lambda})}$ and ${{\mathscr B}({\bf b},\,{\lambda})}$ can be inferred only by calculating the covariance between the corresponding standardized variables ${A({\bf a},\,\lambda)}$ and ${B({\bf b},\,\lambda)}$ as calculated in Ref.${\,}$\cite{IJTP}, namely
\begin{align}
A_{\bf a}({\lambda})\,\equiv\,A({\bf a},\,\lambda)\,&\equiv\,{\bf L}({\bf a},\,\lambda) \label{dumtit-1} \\
\text{and}\;\;\;B_{\bf b}({\lambda})\,\equiv\,B({\bf b},\,\lambda)\,&\equiv\,{\bf L}({\bf b},\,\lambda)\,, \label{dumtit-2}
\end{align}
the correlation between the raw scores ${{\mathscr A}({\bf a},\,{\lambda})}$ and ${{\mathscr B}({\bf b},\,{\lambda})}$ must be obtained by evaluating their product moment
\begin{equation}
{\cal E}({\bf a},\,{\bf b})\,=\lim_{\,n\,\gg\,1}\left[\frac{1}{n}\sum_{k\,=\,1}^{n}\,
{A}({\bf a},\,{\lambda}^k)\;{B}({\bf b},\,{\lambda}^k)\right].\label{stand-exppeu}
\end{equation}
The numerical value of this coefficient is then necessarily equal to the value of the correlation calculated by Eq.${\,}$(\ref{exppeu}). 

Using the above expression for ${{\cal E}({\bf a},\,{\bf b})}$ the string of expectation values (\ref{B1-11}) can now be rewritten as a single average in terms of the standard scores ${A_{\bf a}(\lambda)}$ and ${B_{\bf b}(\lambda)}$, because now they are on equal statistical and geometrical footings:
\begin{equation}
\lim_{\,n\,\gg\,1}\Bigg[\frac{1}{n}\sum_{k\,=\,1}^{n}\,\big\{
A_{\bf a}({\lambda}^k)\,B_{\bf b}({\lambda}^k)\,+\,
A_{\bf a}({\lambda}^k)\,B_{\bf b'}({\lambda}^k)\,+\, A_{\bf a'}({\lambda}^k)\,B_{\bf b}({\lambda}^k)\,-\,
A_{\bf a'}({\lambda}^k)\,B_{\bf b'}({\lambda}^k)\big\}\Bigg]. \label{probnonint}
\end{equation}
But since ${A_{\bf a}({\lambda})\equiv{\bf L}({\bf a},\,\lambda)}$ and ${B_{\bf b}({\lambda})\equiv{\bf L}({\bf b},\,\lambda)}$ are
two independent equatorial points of ${S^3}$, we can take them to belong to two disconnected ``sections'' of ${S^3}$
[{\it i.e.}, two disconnected su(2) 2-spheres within ${S^3\sim {\rm SU}(2)}$], satisfying
\begin{equation}
\left[\,A_{\bf n}({\lambda}),\,B_{\bf n'}({\lambda})\,\right]\,=\,0\,
\;\;\;\forall\;\,{\bf n}\;\,{\rm and}\;\,{\bf n'}\,\in\,{\rm I\!R}^3,\label{com}
\end{equation}
which is equivalent to anticipating a null outcome along the direction ${{\bf n}\times{\bf n'}}$ exclusive to both ${\bf n}$ and ${\bf n'}$. If we now square the integrand of equation (\ref{probnonint}), use the above commutation relations, and use the fact that all unit bivectors square to ${-1}$, then the absolute value of the Bell-CHSH string (\ref{B1-11}) leads to the following variance inequality \cite{Christian}:
\begin{equation}
|{\cal E}({\bf a},\,{\bf b})\,+\,{\cal E}({\bf a},\,{\bf b'})\,+\,
{\cal E}({\bf a'},\,{\bf b})\,-\,{\cal E}({\bf a'},\,{\bf b'})|\,\leqslant\sqrt{\lim_{\,n\,\gg\,1}\left[\frac{1}{n}\sum_{k\,=\,1}^{n}\,
\big\{\,4\,+\,4\,{\mathscr T}_{\,{\bf a\,a'}}({\lambda}^k)\,{\mathscr T}_{\,{\bf b'\,b}}({\lambda}^k)\,\big\}\right]},\label{yever}
\end{equation}
where the classical commutators
\begin{equation}
{\mathscr T}_{\,{\bf a\,a'}}(\lambda):=\frac{1}{2}\left[\,A_{\bf a}(\lambda),\,A_{\bf a'}(\lambda)\right]
\,=\,-\,A_{{\bf a}\times{\bf a'}}(\lambda) \label{aa-potorsion-666}
\end{equation}
and
\begin{equation}
{\mathscr T}_{\,{\bf b'\,b}}(\lambda)
:=\frac{1}{2}\left[\,B_{\bf b'}(\lambda),\,B_{\bf b}(\lambda)\right]\,=\,-\,B_{{\bf b'}\times{\bf b}}(\lambda)\label{bb-potor}
\end{equation}
are the geometric measures of the torsion within ${S^3}$ \cite{Christian}. Thus, as discussed in the paper, it is the non-vanishing torsion
${\mathscr T}$ within the 3-sphere---the parallelizing torsion which makes its Riemann curvature
vanish---that is responsible for the stronger-than-linear correlation. We can see this from Eq.${\,}$(\ref{yever}) by setting
${{\mathscr T}=0}$, and in more detail as follows.

Using definitions (\ref{dumtit-1}) and (\ref{dumtit-2}) for ${A_{\bf a}({\lambda})}$ and
${B_{\bf b}({\lambda})}$ and making a repeated use of the well known bivector identity
\begin{equation}
{\bf L}({\bf a},\,\lambda)\,{\bf L}({\bf a'},\,\lambda)\,=\,-\,{\bf a}\cdot{\bf a'}\,-\,
{\bf L}({\bf a}\times{\bf a'},\,\lambda)\,,\label{bititi}
\end{equation}
the above inequality can be further simplified to
\begin{align}
|{\cal E}({\bf a},\,{\bf b})\,+\,{\cal E}({\bf a},\,{\bf b'})\,+\,
{\cal E}({\bf a'},\,{\bf b})\,-\,{\cal E}({\bf a'},\,{\bf b'})|\,&\leqslant\sqrt{\!4-4\,({{\bf a}}\times{{\bf a}'})\cdot({{\bf b}'}\times{{\bf b}})-
4\!\lim_{\,n\,\gg\,1}\left[\frac{1}{n}\sum_{k\,=\,1}^{n}{{\bf L}}({\bf z},\,\lambda^k)\right]} \notag \\
&\leqslant\sqrt{\!4-4\,({{\bf a}}\times{\bf a'})\cdot({\bf b'}\times{{\bf b}})-
4\!\lim_{\,n\,\gg\,1}\left[\frac{1}{n}\sum_{k\,=\,1}^{n}\lambda^k\right]{{\bf D}}({\bf z})} \notag \\
&\leqslant\,2\,\sqrt{\,1-({{\bf a}}\times{\bf a'})
\cdot({\bf b'}\times{{\bf b}})\,-\,0\,}\,,\label{before-opppo-666}
\end{align}
where ${{\bf z}=({\bf a}\times{\bf a'})\times({\bf b'}\times{\bf b})}$. The last two steps follow from the relation (\ref{OJS}) between ${{\bf L}({\bf z},\,\lambda)}$ and ${{\bf D}({\bf z})}$ and the fact that the orientation ${\lambda}$ of ${S^3}$ is evenly distributed between ${+1}$ and ${-1}$. Finally, by noticing that trigonometry dictates
\begin{equation}
-1\leqslant\,({\bf a}\times{\bf a'})\cdot({\bf b'}\times{\bf b})\,\leqslant +1\,,
\end{equation}
the above inequality can be reduced to the form
\begin{equation}
\left|\,{\cal E}({\bf a},\,{\bf b})\,+\,{\cal E}({\bf a},\,{\bf b'})\,+\,
{\cal E}({\bf a'},\,{\bf b})\,-\,{\cal E}({\bf a'},\,{\bf b'})\,\right|\,\leqslant\,2\sqrt{2}\,,
\label{My-CHSH}
\end{equation}
exhibiting the correct upper bound on the strength of possible correlations. Thus the stronger bounds of ${-2}$ and ${+2}$ calculated na\"ively by Gill in his paper are simply incorrect. More importantly, as correctly done in the paper \cite{IJTP}, the correlation function for the bomb fragments respecting the SU(2) Lie algebra su(2) can indeed be calculated to yield
\begin{equation}
{\cal E}({\bf a},\,{\bf b})=\!\!\lim_{\,n\,\gg\,1}\!\left[\frac{1}{n}\!\sum_{k\,=\,1}^{n}
\{{sign}\,(+{\bf s}^k\cdot{\bf a})\}\,
\{{sign}\,(-{\bf s}^k\cdot{\bf b})\}\right]\!= -\,{\bf a}\cdot{\bf b}\,,\label{correlations}
\end{equation}
where ${n}$ is the total number of trials performed. Therefore the remaining comments by Gill are anything but justified.

In fact there is considerable confusion in Gill's attempted misrepresentation of the proposed experiment \cite{Gillprint}. Rather surprisingly, his comments fail to distinguish between the {\it theoretical} derivation of the correlation presented in Eq.${\,}$(110) of Ref.${\,}$\cite{IJTP} and the {\it practical} calculation of the correlation expressed in the above equation in terms of the raw scores. For example, he asserts that ``correlations should only be computed the usual way using actual experimental outcomes..." This reveals that Gill has either not bothered to read the experimental proposal discussed in paper \cite{IJTP}, or has failed to understand the crucial difference between the traditional Bell-type experiments and the one proposed in the paper.

The important question here is: What would be the ``actual experimental outcomes" in the proposed experiment? From the above discussion and the discussion in the Section 4 of Ref.${\,}$\cite{IJTP} it is quite clear that what would be actually observed in the experiment are the bivectors ${{\bf L}({\bf a},\,\lambda)}$ and ${{\bf L}({\bf b},\,\lambda)}$ about the directions ${\bf a}$ and ${\bf b}$, respectively. The raw scores ${{sign}\,(\,+\,{\bf s}^k\cdot{\bf a})}$ and ${{sign}\,(\,-\,{\bf s}^k\cdot{\bf b})}$ may then be calculated, but by an algorithm {\it extraneous} to the actual experiment, and {\it long after} the experiment has actually been completed. Thus they would not be an integral dynamical part of the physical experiment itself. On the other hand, thanks to the precise geometrical relation between the raw scores ${{sign}\,(\,+\,{\bf s}^k\cdot{\bf a})}$ and ${{sign}\,(\,-\,{\bf s}^k\cdot{\bf b})}$ and the standard scores ${{\bf L}({\bf a},\,\lambda)}$ and ${{\bf L}({\bf b},\,\lambda)}$ given by the equations (\ref{88-oi}) and (\ref{99-oi}), the statistical association between the raw scores can nevertheless be inferred by calculating the {\it covariance} of the corresponding standardized variables ${A_{\bf a}(\lambda)}$ and ${B_{\bf b}(\lambda)}$ as demonstrated above, thus predicting the correlation (\ref{correlations}).

Finally, Gill falsely and disingenuously claims that Section 5 of Ref.${\,}$\cite{IJTP} reproduces a sign error which is also present in my earlier work \cite{Christian}. There is in fact no such error, as explained already in Ref.${\,}$\cite{refute} and in several chapters of Ref.${\,}$\cite{Christian}.

In conclusion, the criticism of the proposed experiment by Gill is a travesty. No physicist should be deceived by it.

\appendix
\section{Refutations of Gill's Mistaken Claims by Explicit Numerical Simulations}

The above refutations of Gill's fallacious claims have been independently verified by Albert Jan Wonnink \cite{Wonnink} in an explicit numerical simulation of Eq.${\,}$(\ref{stand-exppeu}), by means of a specialized program for geometric algebra based computations. In other words, Gill's mistaken claims have been refuted by Wonnink by numerically computing the expectation value 
\begin{equation}
{\cal E}({\bf a},\,{\bf b})\,=\lim_{\,n\,\gg\,1}\left[\frac{1}{n}\sum_{k\,=\,1}^{n}\,
{\bf L}({\bf a},\,{\lambda}^k)\;{\bf L}({\bf b},\,{\lambda}^k)\right]=-{\bf a}\cdot{\bf b}.\label{stand-exsss}
\end{equation}
To understand this computation, recall that ${{\bf L}({\bf a},\,{\lambda}^k=+1)=+\,I\cdot{\bf a}}$ and ${{\bf L}({\bf b},\,{\lambda}^k=-1)=-\,I\cdot{\bf b}}$ represent the two spins of the bomb fragments, where ${I:=e_x\wedge e_y\wedge e_z}$ with ${+I}$ representing the right-handed orientation of ${S^3}$ and ${-I}$ representing the left-handed orientation of ${S^3}$. Consequently we may consider the following two geometric products, 
\begin{equation}
(\,+\,I\cdot{\bf a})(\,+\,I\cdot{\bf b})\,
=\,-\,{\bf a}\cdot{\bf b}\,-\,(\,+\,I\,)\cdot({\bf a}\times{\bf b}) \label{id-1}
\end{equation}
and
\begin{equation}
(\,-\,I\cdot{\bf a})(\,-\,I\cdot{\bf b})\,
=\,-\,{\bf a}\cdot{\bf b}\,-\,(\,-\,I\,)\cdot({\bf a}\times{\bf b}), \label{id-2}
\end{equation}
as discussed in my earlier reply to Gill (cf. Eqs.${\,}$(15) and (16) of Ref.${}$\cite{refute}). The two possible orientations of ${S^3}$ may then be thought of as 
the random hidden variables ${\lambda=\pm\,1}$ (or the initial states ${\lambda=\pm\,1}$) of the two bomb fragments.

Now in traditional geometric algebra as well as in the GAViewer program \cite{GAV} employed by Albert Jan Wonnink the volume form of the physical space is fixed {\it a priori} to be ${+I}$, by convention. This convention is inconsistent with the physical process of spin detections delineated in the Eqs.${\,}$(\ref{88-oi}) and (\ref{99-oi}) above. Therefore a translation of the geometric product ${{\bf L}({\bf a},\,{\lambda}^k=-1)\;{\bf L}({\bf b},\,{\lambda}^k=-1)=(\,-\,I\cdot{\bf a})(\,-\,I\cdot{\bf b})}$ for the GAViewer built on the right-handed form ${+I}$ is necessary, which can be inferred from Eqs.${\,}$(\ref{id-1}) and (\ref{id-2}) by recalling that ${{\bf b}\times{\bf a}=-\,{\bf a}\times{\bf b}}$ and rewriting Eq.${\,}$(\ref{id-1}) as
\begin{equation}
(\,+\,I\cdot{\bf b})(\,+\,I\cdot{\bf a})\,
=\,-\,{\bf a}\cdot{\bf b}\,+\,(\,+\,I\,)\cdot({\bf a}\times{\bf b})\,=\,-\,{\bf a}\cdot{\bf b}\,-\,(\,-\,I\,)\cdot({\bf a}\times{\bf b}). \label{id-3}
\end{equation}
Comparing the right-hand sides of Eqs.${\,}$(\ref{id-2}) and (\ref{id-3}) it is now quite easy to recognize that the desired translation is 
\begin{equation}
(\,-\,I\cdot{\bf a})(\,-\,I\cdot{\bf b})\,\longrightarrow\,(\,+\,I\cdot{\bf b})(\,+\,I\cdot{\bf a}). \label{traa}
\end{equation}
In terms of this translation the geometric products appearing in the expectation function (\ref{stand-exsss}) can be expressed as
\begin{align}
{\bf L}({\bf a},\,{\lambda}^k=+1)\;{\bf L}({\bf b},\,{\lambda}^k=+1)\,&=\,(\,+\,I\cdot{\bf a})(\,+\,I\cdot{\bf b}) \\
\text{and}\;\;\;{\bf L}({\bf a},\,{\lambda}^k=-1)\;{\bf L}({\bf b},\,{\lambda}^k=-1)\,&=\,(\,+\,I\cdot{\bf b})(\,+\,I\cdot{\bf a}).
\end{align}
In other words, when ${\lambda^k}$ happens to be equal to ${+1}$, ${{\bf L}({\bf a},\,{\lambda}^k)\;{\bf L}({\bf b},\,{\lambda}^k)=(\,+\,I\cdot{\bf a})(\,+\,I\cdot{\bf b})}$, and when ${\lambda^k}$ happens to be equal to ${-1}$, ${{\bf L}({\bf a},\,{\lambda}^k)\;{\bf L}({\bf b},\,{\lambda}^k)=(\,+\,I\cdot{\bf b})(\,+\,I\cdot{\bf a})}$. Consequently, the expectation value (\ref{stand-exsss}) reduces at once to
\begin{equation}
{\cal E}({\bf a},\,{\bf b})\,=\,\frac{1}{2}(\,+\,I\cdot{\bf a})(\,+\,I\cdot{\bf b})\,+\,\frac{1}{2}(\,+\,I\cdot{\bf b})(\,+\,I\cdot{\bf a})\,
=\,-\,{\bf a}\cdot{\bf b}\,+\,0\,,\label{stand-nossss}
\end{equation}
because the orientation ${\lambda}$ of ${S^3}$ is necessarily a fair coin. Here the last equality follows from the Eqs.${\,}$(\ref{id-1}) and (\ref{id-3}).

Given the translation (\ref{traa}), it is now easy to understand the simulation code of Ref.${\,}$\cite{Wonnink}, with its essential lines being 
\begin{align}
\text{if}\;\;\lambda\,&=\,+1,\;\;\text{then add }\;\;(\,+\,I\cdot{\bf a})(\,+\,I\cdot{\bf b}), \\
\text{but if}\;\;\lambda\,&=\,-1,\;\;\text{then add }\;\;(\,+\,I\cdot{\bf b})(\,+\,I\cdot{\bf a}).
\end{align}
Not surprisingly, the expectation value computed in this event-by-event simulation prints out to be ${{\cal E}({\bf a},\,{\bf b})=-\,{\bf a}\cdot{\bf b}}$.

In complement to the above unambiguous demonstration, it is also possible to verify the correlation (\ref{stand-nossss}) numerically in an event-by-event simulation within a {\it non}-Clifford algebraic representation of ${S^3}$, as done, for example, in Ref.${\,}$\cite{rpub}.

Thus, once again, the criticism of the proposed experiment by Gill is exposed to be vacuous by explicit computations.

\section{Correlations Among the Raw Scores is Equal to the Covariance Among the Standard Scores}

For the completeness of the arguments presented above, in this appendix let us prove the following equality explicitly: 
\begin{equation}
{\cal E}({\bf a},\,{\bf b})\,=\!\lim_{\,n\,\gg\,1}\left[\frac{1}{n}\!\sum_{k\,=\,1}^{n}
\{{sign}\,(+{\bf s}^k\cdot{\bf a})\}\,
\{{sign}\,(-{\bf s}^k\cdot{\bf b})\}\right]=\!\lim_{\,n\,\gg\,1}\left[\frac{1}{n}\sum_{k\,=\,1}^{n}\,
{\bf L}({\bf a},\,{\lambda}^k)\;{\bf L}({\bf b},\,{\lambda}^k)\right]=-{\bf a}\cdot{\bf b}.\label{nexs-bss}
\end{equation}
We begin with the central equalities in the prescription of the remote measurement events given in Eqs.${\,}$(\ref{88-oi}) and (\ref{99-oi}):
\begin{align}
{sign}\,(\,+\,{\bf s}^k\cdot{\bf a})\,=\,{\mathscr A}({\bf a},\,{\lambda^k})\,&=\,\lim_{{\bf s}\,\rightarrow\,{\bf a}}\left\{\,-\,{\bf D}({\bf a})\,{\bf L}({\bf s},\,\lambda^k)\,\right\} \label{ppp-oi}\\
\text{and}\;\;{sign}\,(\,-\,{\bf s}^k\cdot{\bf b})\,=\,{\mathscr B}({\bf b},\,{\lambda^k})\,&=\,\lim_{{\bf s}\,\rightarrow\,{\bf b}}\left\{\,+\,{\bf D}({\bf b})\,{\bf L}({\bf s},\,\lambda^k)\,\right\}, \label{000-oi}
\end{align}
where the {\it same} scalar number ${{\mathscr A}({\bf a},\,{\lambda^k})=\pm1}$ is expressed in two different ways --- as a grade-0 number ${{sign}\,(\,+\,{\bf s}^k\cdot{\bf a})}$ on the left hand, and as a product ${-\,{\bf D}({\bf a})\,{\bf L}({\bf a},\,\lambda^k)}$ of two grade-2 numbers, ${-\,{\bf D}({\bf a})}$ and ${{\bf L}({\bf a},\,\lambda^k)}$, on the right hand.

Now the correlation between ${{\mathscr A}({\bf a},\,{\lambda^k})}$ and ${{\mathscr B}({\bf b},\,{\lambda^k})}$ can be quantified by the product-moment correlation coefficient
\begin{equation}
{\cal E}({\bf a},\,{\bf b})\,=\,\frac{{\displaystyle\lim_{\,n\,\gg\,1}}\left[{\displaystyle\frac{1}{n}\sum_{k\,=\,1}^{n}}\,
\left\{{\mathscr A}({\bf a},\,{\lambda}^k)-\overline{{\mathscr A}({\bf a},\,{\lambda})}\right\}\;\left\{{\mathscr B}({\bf b},\,{\lambda}^k)-\overline{{\mathscr B}({\bf b},\,{\lambda})}\right\}\right]}{\sigma({\mathscr A})\;\sigma({\mathscr B})}\,,\label{exs-bss}
\end{equation}
where ${\overline{{\mathscr A}({\bf a},\,{\lambda})}}$ and ${\overline{{\mathscr B}({\bf b},\,{\lambda})}}$ are the average values of ${\mathscr A}$ and ${\mathscr B}$ and ${\sigma({\mathscr A})}$ and ${\sigma({\mathscr B})}$ are their standard deviations:
\begin{equation}
\sigma({\mathscr A})\,=\,\sqrt{\frac{1}{n}\sum_{k\,=\,1}^{n}\,\left|\left|\,{\mathscr A}({\bf a},\,{\lambda}^k)\,-\,
{\overline{{\mathscr A}({\bf a},\,{\lambda})}}\;\right|\right|^2\,}\, \;\;\;\;\;\text{and}\;\;\;\;\;
\sigma({\mathscr B})\,=\,\sqrt{\frac{1}{n}\sum_{k\,=\,1}^{n}\,\left|\left|\,{\mathscr B}({\bf b},\,{\lambda}^k)\,-\,{\overline{{\mathscr B}({\bf b},\,{\lambda})}}\;\right|\right|^2\,}.
\label{defstan}
\end{equation}
Accordingly, let us first consider ${{\mathscr A}({\bf a},\,{\lambda^k})={sign}\,(\,+\,{\bf s}^k\cdot{\bf a})}$ and ${{\mathscr B}({\bf b},\,{\lambda^k})={sign}\,(\,-\,{\bf s}^k\cdot{\bf b})}$ from the left sides of the equalities (\ref{ppp-oi}) and (\ref{000-oi}). Written in this form, these variables cannot be factorized into products of a random number and a non-random number. We are therefore forced to treat them as irreducible random variables. Moreover, since ${\lambda}$ is a fair coin, ${\bf s}$ has equal chance of being parallel and anti-parallel to ${\bf a}$. Consequently, it is easy to work out from the above formulae that in the present case ${\overline{{\mathscr A}({\bf a},\,{\lambda})}=\overline{{\mathscr B}({\bf b},\,{\lambda})}=0\,}$ and ${\,\sigma({\mathscr A})=\sigma({\mathscr B})=1}$, which immediately gives us
\begin{equation}
{\cal E}({\bf a},\,{\bf b})\,=\!\lim_{\,n\,\gg\,1}\!\left[\frac{1}{n}\!\sum_{k\,=\,1}^{n}
\{{sign}\,(+{\bf s}^k\cdot{\bf a})\}\,
\{{sign}\,(-{\bf s}^k\cdot{\bf b})\}\right]\!.\label{pers-bss}
\end{equation}
Thus, we recognize that the above expression of the observed correlations --- which is traditionally employed by the experimentalists --- is simply a special case of the Pearson's product-moment correlation coefficient defined in Eq.${\,}$(\ref{exs-bss}). 

Next, consider ${{\mathscr A}({\bf a},\,{\lambda^k})=-\,{\bf D}({\bf a})\,{\bf L}({\bf a},\,\lambda^k)}$ and ${{\mathscr B}({\bf b},\,{\lambda^k})=+\,{\bf D}({\bf b})\,{\bf L}({\bf b},\,\lambda^k)}$ from the right sides of the equalities (\ref{ppp-oi}) and (\ref{000-oi}). Written in this form, these very same variables are now factorized into geometric products of a random number and a non-random number. Evidently, the variable ${{\mathscr A}({\bf a},\,{\lambda}^k)}$ is a product of a random spin bivector ${{\bf L}({\bf a},\,\lambda^k)}$ (which is a function of the random variable ${\lambda}$) and a non-random detector bivector ${-\,{\bf D}({\bf a})}$ (which is independent of the random variable ${\lambda}$). In other words, the randomness within ${{\mathscr A}({\bf a},\,{\lambda^k})}$ originates entirely from the randomness of the spin bivector ${{\bf L}({\bf a},\,\lambda^k)}$, with the detector bivector ${-\,{\bf D}({\bf a})}$ merely specifying the scale of dispersion within ${{\bf L}({\bf a},\,\lambda^k)}$. Consequently, as random variables, ${{\mathscr A}({\bf a},\,{\lambda}^k)}$ and ${{\mathscr B}({\bf b},\,{\lambda}^k)}$ are generated with {\it different} standard deviations, or {\it different} sizes of the typical error. ${{\mathscr A}({\bf a},\,{\lambda}^k)}$ is generated with a typical error ${-\,{\bf D}({\bf a})}$, whereas ${{\mathscr B}({\bf b},\,{\lambda}^k)}$ is generated with a typical error ${+\,{\bf D}({\bf b})}$. And since errors in linear relations propagate linearly, the standard deviation ${\sigma({\mathscr A}\,)}$ of ${{\mathscr A}({\bf a},\,{\lambda}^k)}$ is equal to ${-\,{\bf D}({\bf a})}$ times the standard deviation of ${{\bf L}({\bf a},\,\lambda^k)}$, and the standard deviation ${\sigma({\mathscr B}\,)}$ of ${{\mathscr B}({\bf b},\,{\lambda}^k)}$ is equal to ${+\,{\bf D}({\bf b})}$ times the standard deviation of ${{\bf L}({\bf b},\,\lambda^k)}$, which can be worked out using formulae similar to (\ref{defstan}):
\begin{align}
\sigma({\mathscr A}\,)\,=\,-\,{\bf D}({\bf a})\;\sigma\!\left\{{\bf L}({\bf a},\,\lambda^k)\right\}\,&=\,-\,{\bf D}({\bf a})\,\sqrt{\frac{1}{n}\sum_{k\,=\,1}^{n}\,\left|\left|\,{\bf L}({\bf a},\,{\lambda}^k)\,-\,
{\overline{{\bf L}({\bf a},\,{\lambda})}}\;\right|\right|^2\,}\,=\,-\,{\bf D}({\bf a})\\
\text{and}\;\,\;\sigma({\mathscr B}\,)\,=\,+\,{\bf D}({\bf b})\;\sigma\!\left\{{\bf L}({\bf b},\,\lambda^k)\right\}\,&=\,+\,{\bf D}({\bf b})\,\sqrt{\frac{1}{n}\sum_{k\,=\,1}^{n}\,\left|\left|\,{\bf L}({\bf b},\,{\lambda}^k)\,-\,
{\overline{{\bf L}({\bf b},\,{\lambda})}}\;\right|\right|^2\,}\,=\,+\,{\bf D}({\bf b}).\;\,\text{}
\end{align}
Here I have used ${\left|\left|\,{\bf L}({\bf a},\,\lambda^k)\,\right|\right|=\left|\left|\,{\bf L}({\bf b},\,\lambda^k)\,\right|\right|=1}$ because all ${{\bf L}({\bf s},\,\lambda^k)}$ are {\it unit} bivectors, and ${{\overline{{\bf L}({\bf a},\,{\lambda})}}={\overline{{\bf L}({\bf b},\,{\lambda})}}=0}$ on the account of ${\lambda}$ being a fair coin. These standard deviations are derived much more rigorously in Refs.${\,}$\cite{IJTP} and \cite{Christian}. 

If we now substitute ${{\mathscr A}({\bf a},\,{\lambda^k})=-\,{\bf D}({\bf a})\,{\bf L}({\bf a},\,\lambda^k)}$ and ${{\mathscr B}({\bf b},\,{\lambda^k})=+\,{\bf D}({\bf b})\,{\bf L}({\bf b},\,\lambda^k)}$ into the same formula (\ref{exs-bss}) used previously for the product-moment correlation coefficient \cite{refute}, together with ${\overline{{\mathscr A}({\bf a},\,{\lambda})}=\overline{{\mathscr B}({\bf b},\,{\lambda})}=0}$, ${\sigma({\mathscr A}\,)=-\,{\bf D}({\bf a})}$, ${\sigma({\mathscr B}\,)=+\,{\bf D}({\bf b})}$, ${\{{\mathscr A}({\bf a},\,{\lambda^k})/{\sigma({\mathscr A}\,)}\}={\bf L}({\bf a},\,\lambda^k)}$, and ${\{{\mathscr B}({\bf b},\,{\lambda^k})/{\sigma({\mathscr B}\,)}\}={\bf L}({\bf b},\,\lambda^k)}$, then it immediately reduces to
\begin{equation}
{\cal E}({\bf a},\,{\bf b})\,=\lim_{\,n\,\gg\,1}\left[\frac{1}{n}\sum_{k\,=\,1}^{n}\,
{\bf L}({\bf a},\,{\lambda}^k)\;{\bf L}({\bf b},\,{\lambda}^k)\right]\!.\label{tands-exsss}
\end{equation}
Consequently, putting the results (\ref{pers-bss}), (\ref{tands-exsss}), and (\ref{stand-nossss}) together, we finally arrive at the equality we set out to prove: 
\begin{equation}
{\cal E}({\bf a},\,{\bf b})\,=\!\lim_{\,n\,\gg\,1}\left[\frac{1}{n}\!\sum_{k\,=\,1}^{n}
\{{sign}\,(+{\bf s}^k\cdot{\bf a})\}\,
\{{sign}\,(-{\bf s}^k\cdot{\bf b})\}\right]=\!\lim_{\,n\,\gg\,1}\left[\frac{1}{n}\sum_{k\,=\,1}^{n}\,
{\bf L}({\bf a},\,{\lambda}^k)\;{\bf L}({\bf b},\,{\lambda}^k)\right]=-{\bf a}\cdot{\bf b}.\label{meexs-bss}
\end{equation}

\section{A Simplified Local-Realistic Derivation of the EPR-Bohm Correlation {\rm\cite{disproof}}}

As in Eqs.${\,}$(\ref{88-oi}) and (\ref{99-oi}), let the spin bivectors ${\mp\,{\bf L}({\bf s},\,\lambda^k)}$ be detected by the detector bivectors ${{\bf D}({\bf a})}$ and ${{\bf D}({\bf b})}$, giving
\begin{align}
S^3\ni\,{\mathscr A}({\bf a},\,{\lambda^k})\,:=\,\lim_{{\bf s}\,\rightarrow\,{\bf a}}\left\{-\,{\bf D}({\bf a})\,{\bf L}({\bf s},\,\lambda^k)\right\}&=\,                  
\begin{cases}
+\,1\;\;\;\;\;{\rm if} &\lambda^k\,=\,+\,1 \\
-\,1\;\;\;\;\;{\rm if} &\lambda^k\,=\,-\,1
\end{cases} \Bigg\}\,\;\text{with}\;\, \Bigl\langle\,{\mathscr A}({\bf a},\,\lambda^k)\,\Bigr\rangle\,=\,0\\
\text{and}\;\;\;\;S^3\ni\,{\mathscr B}({\bf b},\,{\lambda^k})\,:=\,\lim_{{\bf s}\,\rightarrow\,{\bf b}}\left\{+\,{\bf L}({\bf s},\,\lambda^k)\,{\bf D}({\bf b})\right\}&=\,                     
\begin{cases}
-\,1\;\;\;\;\;{\rm if} &\lambda^k\,=\,+\,1 \\
+\,1\;\;\;\;\;{\rm if} &\lambda^k\,=\,-\,1
\end{cases} \Bigg\}\,\;\text{with}\;\,\Bigl\langle\,{\mathscr B}({\bf b},\,\lambda^k)\,\Bigr\rangle\,=\,0\,,
\label{aaa99-oi}
\end{align}
where the orientation ${\lambda}$ of ${S^3}$ is assumed to be a random variable with 50/50 chance of being ${+1}$ or ${-\,1}$ at the moment of the pair-creation, making the spinning bivector ${{\bf L}({\bf n},\,\lambda^k)}$ a random variable {\it relative} to the detector bivector ${{\bf D}({\bf n})}$:
\begin{equation}
{\bf L}({\bf n},\,\lambda^k)\,=\,\lambda^k\,{\bf D}({\bf n})\,\,\Longleftrightarrow\,\,{\bf D}({\bf n})\,=\,\lambda^k\,{\bf L}({\bf n},\,\lambda^k)\,. \label{aaaOJS}
\end{equation}
The expectation value of simultaneous outcomes ${{\mathscr A}({\bf a},\,{\lambda^k})=\pm1}$ and ${{\mathscr B}({\bf b},\,{\lambda^k})=\pm1}$ in ${S^3}$ then works out as follows: 
\begin{align}
{\cal E}({\bf a},\,{\bf b})\,&=\lim_{\,n\,\rightarrow\,\infty}\left[\frac{1}{n}\sum_{k\,=\,1}^{n}\,
{\mathscr A}({\bf a},\,{\lambda}^k)\;{\mathscr B}({\bf b},\,{\lambda}^k)\right]\,\text{within}\,\;S^3:=\,\text{the set of all unit (left-handed) quaternions \cite{IJTP}\cite{Christian}} \\
&=\lim_{\,n\,\rightarrow\,\infty}\left[\frac{1}{n}\sum_{k\,=\,1}^{n}\,\bigg[\lim_{{\bf s}\,\rightarrow\,{\bf a}}\left\{\,-\,{\bf D}({\bf a})\,{\bf L}({\bf s},\,\lambda^k)\right\}\bigg]\left[\lim_{{\bf s}\,\rightarrow\,{\bf b}}\left\{\,+\,{\bf L}({\bf s},\,\lambda^k)\,{\bf D}({\bf b})\right\}\,\right]\right]\;\;\text{(conserving total spin = 0)} \\
&=\lim_{\,n\,\rightarrow\,\infty}\left[\frac{1}{n}\sum_{k\,=\,1}^{n}\,\lim_{\substack{{\bf s}\,\rightarrow\,{\bf a} \\ {\bf s}\,\rightarrow\,{\bf b}}}\left\{\,-\,{\bf D}({\bf a})\,{\bf L}({\bf s},\,\lambda^k)\,\,{\bf L}({\bf s},\,\lambda^k)\,{\bf D}({\bf b})\,\equiv\,{\bf q}({\bf a},\,{\bf b};\,{\bf s},\,\lambda^k)\right\}\right]\;\;\,\text{(by a property of limits)} \\
&=\lim_{\,n\,\rightarrow\,\infty}\left[\frac{1}{n}\sum_{k\,=\,1}^{n}\,\lim_{\substack{{\bf s}\,\rightarrow\,{\bf a} \\ {\bf s}\,\rightarrow\,{\bf b}}}\left\{\,-\,\lambda^k\,{\bf L}({\bf a},\,\lambda^k)\,\,{\bf L}({\bf s},\,\lambda^k)\,{\bf L}({\bf s},\,\lambda^k)\,\,\lambda^k\,{\bf L}({\bf b},\,\lambda^k)\right\}\right]\text{(all bivectors in the spin basis)}\\
&=\lim_{\,n\,\rightarrow\,\infty}\left[\frac{1}{n}\sum_{k\,=\,1}^{n}\,\lim_{\substack{{\bf s}\,\rightarrow\,{\bf a} \\ {\bf s}\,\rightarrow\,{\bf b}}}\left\{\,-\,{\bf L}({\bf a},\,\lambda^k)\,\,{\bf L}({\bf s},\,\lambda^k)\,{\bf L}({\bf s},\,\lambda^k)\,\,{\bf L}({\bf b},\,\lambda^k)\right\}\right]\,\text{(scalars ${\lambda^k}$ commute with bivectors)}\\
&=\lim_{\,n\,\rightarrow\,\infty}\left[\frac{1}{n}\sum_{k\,=\,1}^{n}\,{\bf L}({\bf a},\,\lambda^k)\,{\bf L}({\bf b},\,\lambda^k)\,\right]\,\text{(follows from the conservation of zero spin angular momentum)} \\
&=\,-\,{\bf a}\cdot{\bf b}\,-\!\lim_{\,n\,\rightarrow\,\infty}\left[\frac{1}{n}\sum_{k\,=\,1}^{n}\,{\bf L}({\bf a}\times{\bf b},\,\lambda^k)\,\right]\;\;\,\text{(NB: there is no ``third" spin about the direction ${{\bf a}\times{\bf b}}$)} \\
&=\,-\,{\bf a}\cdot{\bf b}\,-\!\lim_{\,n\,\rightarrow\,\infty}\left[\frac{1}{n}\sum_{k\,=\,1}^{n}\,\lambda^k\,\right]{\bf D}({\bf a}\times{\bf b})\;\;\text{(summing over counterfactual detections of ``third" spins)} \\
&=\,-\,{\bf a}\cdot{\bf b}\,+\,0\,\;\text{(because the scalar coefficient of the bivector ${{\bf D}({\bf a}\times{\bf b})}$ vanishes in the ${n\rightarrow\infty}$ limit)}
\label{exppeuuu}
\end{align}
Here the integrand of (C6) is necessarily a unit quaternion ${{\bf q}({\bf a},\,{\bf b};\,{\bf s},\,\lambda^k)}$ since ${S^3}$ remains closed under multiplication; (C7) follows from using (C3); (C8) follows from using ${\lambda^2 = +1}$; (C9) follows from the fact that all unit bivectors such\break as ${{\bf L}({\bf s},\,\lambda^k)}$ square to ${-1}$; (C10) follows from the geometric product (\ref{bititi}); (C11) follows from using (C3); (C12) follows from the fact that initial orientation ${\lambda}$ of ${S^3}$ is a fair coin; and (C6) follows from (C5) as a special case of the identity 
\begin{equation}
\bigg[\lim_{{\bf s}\,\rightarrow\,{\bf a'}}\left\{\,-\,{\bf D}({\bf a})\,{\bf L}({\bf s},\,\lambda^k)\right\}\bigg]\left[\lim_{{\bf s}\,\rightarrow\,{\bf b'}}\left\{\,+\,{\bf L}({\bf s},\,\lambda^k)\,{\bf D}({\bf b})\right\}\,\right]
=\lim_{\substack{{\bf s}\,\rightarrow\,{\bf a'} \\ {\bf s}\,\rightarrow\,{\bf b'}}}\bigg\{-\,{\bf D}({\bf a})\,{\bf L}({\bf s},\,\lambda^k)\,{\bf L}({\bf s},\,\lambda^k)\,{\bf D}({\bf b})\,\bigg\},
\end{equation}
which can be easily verified either by immediate inspection or by recalling the elementary properties of limits. Note that\break apart from the assumption (C3) of initial state ${\lambda}$ the only other assumption needed in this derivation is the conservation of zero spin angular momentum. These two assumptions are necessary and sufficient to dictate the singlet correlation:      
\begin{align}
{\cal E}({\bf a},\,{\bf b})\,&=\lim_{\,n\,\rightarrow\,\infty}\left[\frac{1}{n}\sum_{k\,=\,1}^{n}\,
{\mathscr A}({\bf a},\,{\lambda}^k)\;{\mathscr B}({\bf b},\,{\lambda}^k)\right]
=\,-\,{\bf a}\cdot{\bf b}\,.
\end{align}
This demonstrates that EPR-Bohm correlations are correlations among the scalar points of a quaternionic 3-sphere.

\section{On the Fatal Mistake Made by John S. Bell in the Proof of His Famous Theorem}

\parskip 13pt
\baselineskip 15pt

Consider -- in slightly modern terms -- the standard EPR-Bohm type experiment envisaged by John S. Bell to prove his famous theorem \cite{Bell}. Alice is free to choose a detector direction ${\bf a}$ or ${\bf a'}$ and Bob is free to choose a detector direction ${\bf b}$ or ${\bf b'}$ to detect spins of the fermions they receive from a common source, at a space-like distance from each other. The objects of interest then are the bounds on the sum of possible averages put together in the manner of CHSH \cite{CHSH},
\begin{equation}
{\cal E}({\bf a},\,{\bf b})\,+\,{\cal E}({\bf a},\,{\bf b'})\,+\,{\cal E}({\bf a'},\,{\bf b})\,-\,{\cal E}({\bf a'},\,{\bf b'})\,, \label{B1-11-2}
\end{equation}
with each average defined as
\begin{equation}
{\cal E}({\bf a},\,{\bf b})\,=\lim_{\,n\,\gg\,1}\left[\frac{1}{n}\sum_{k\,=\,1}^{n}\,
{\mathscr A}({\bf a},\,{\lambda}^k)\;{\mathscr B}({\bf b},\,{\lambda}^k)\right]\,\equiv\,\Bigl\langle\,{\mathscr A}_{k}({\bf a})\,{\mathscr B}_{k}({\bf b})\,\Bigr\rangle\,,\label{exppeu-2}
\end{equation}
where ${\mathscr A({\bf a},\,{\lambda}^k)\equiv {\mathscr A}_{k}({\bf a})=\pm1}$ and ${\mathscr B({\bf b},\,{\lambda}^k)\equiv {\mathscr B}_{k}({\bf b})=\pm1}$ are the respective measurement results of Alice and Bob. Now, since ${{\mathscr A}_{k}({\bf a})=\pm1}$ and ${{\mathscr B}_{k}({\bf b})=\pm1}$, the average of their product is ${-1\leqslant\Bigl\langle\,{\mathscr A}_{k}({\bf a})\,{\mathscr B}_{k}({\bf b})\,\Bigr\rangle\leqslant +1}$. As a result, we can immediately read off the upper and lower bounds on the string of the four averages considered above in (\ref{B1-11-2}):
\begin{equation}
-\,4\,\leqslant\,\Bigl\langle\,{\mathscr A}_{k}({\bf a})\,{\mathscr B}_{k}({\bf b})\,\Bigr\rangle\,+\, \Bigl\langle\,{\mathscr A}_{k}({\bf a})\,{\mathscr B}_{k}({\bf b'})\,\Bigr\rangle\,+\,\Bigl\langle\,{\mathscr A}_{k}({\bf a'})\,{\mathscr B}_{k}({\bf b})\,\Bigr\rangle\,-\, \Bigl\langle\,{\mathscr A}_{k}({\bf a'})\,{\mathscr B}_{k}({\bf b'})\,\Bigr\rangle\,\leqslant\,+\,4\,. \label{3}
\end{equation}

This should have been the final conclusion by Bell. But he continued. And in doing so he made one of his gravest mistakes. He replaced the above string of four separate averages of binary numbers with the following single average:
\begin{equation}
{\cal E}({\bf a},\,{\bf b})\,+\,{\cal E}({\bf a},\,{\bf b'})\,+\,{\cal E}({\bf a'},\,{\bf b})\,-\,{\cal E}({\bf a'},\,{\bf b'})\,\longrightarrow\,\Bigl\langle\,{\mathscr A}_{k}({\bf a})\,{\mathscr B}_{k}({\bf b})\,+\,
{\mathscr A}_{k}({\bf a})\,{\mathscr B}_{k}({\bf b'})\,+\,{\mathscr A}_{k}({\bf a'})\,{\mathscr B}_{k}({\bf b})\,-\,
{\mathscr A}_{k}({\bf a'})\,{\mathscr B}_{k}({\bf b'})\,\Bigr\rangle\,. \label{rep}
\end{equation}
As innocuous as this step may seem, it is in fact an illegitimate step, for what is being averaged on the RHS are {\it un-observable}, and hence {\it un-physical} quantities. But this illegitimate step allows one to reduce the above average to 
\begin{equation}
\Bigl\langle\,{\mathscr A}_{k}({\bf a})\,\big\{\,{\mathscr B}_{k}({\bf b})+{\mathscr B}_{k}({\bf b'})\,\big\}\,+\,{\mathscr A}_{k}({\bf a'})\,\big\{\,{\mathscr B}_{k}({\bf b})-{\mathscr B}_{k}({\bf b'})\,\big\}\,\Bigr\rangle\,. \label{5}
\end{equation}
And since ${{\mathscr B}_{k}({\bf b})=\pm1}$, if ${|{\mathscr B}_{k}({\bf b})+{\mathscr B}_{k}({\bf b'})|=2}$, then ${|{\mathscr B}_{k}({\bf b})-{\mathscr B}_{k}({\bf b'})|=0}$, and vice versa. Consequently, using ${{\mathscr A}_{k}({\bf a})=\pm1}$, it is easy to conclude that the absolute value of the above average cannot exceed 2, just as Bell concluded: 
\begin{equation}
-\,2\,\leqslant\,\Bigl\langle\,{\mathscr A}_{k}({\bf a})\,{\mathscr B}_{k}({\bf b})\,+\,
{\mathscr A}_{k}({\bf a})\,{\mathscr B}_{k}({\bf b'})\,+\,{\mathscr A}_{k}({\bf a'})\,{\mathscr B}_{k}({\bf b})\,-\,
{\mathscr A}_{k}({\bf a'})\,{\mathscr B}_{k}({\bf b'})\,\Bigr\rangle\,\leqslant\,+\,2\,.
\end{equation}

Let us now try to understand why the replacement (\ref{rep}) is illegal. To begin with, Einstein's (or even Bell's own) notion of local-realism does not demand this replacement. Since this notion is captured already in the definition of the measurement functions ${{\mathscr A}({\bf a},\,{\lambda}^k)}$, the LHS of (\ref{rep}) satisfies the demand of local-realism perfectly well. To be sure, mathematically there is nothing wrong with a replacement of four separate averages with a single average. Every school child knows that the sum of averages is equal to the average of sums. But this rule of thumb is not valid in the above case, because ${({\bf a},\,{\bf b})}$, ${({\bf a},\,{\bf b'})}$, ${({\bf a'},\,{\bf b})}$, and ${({\bf a'},\,{\bf b'})}$ are {\it mutually exclusive pairs of measurement directions}, corresponding to four {\it incompatible} experiments. Each pair can be used by Alice and Bob for a given experiment, for all runs ${1}$ to ${n}$, but no two of the four pairs can be used by them simultaneously. This is because Alice and Bob do not have the ability to make measurements along counterfactually possible pairs of directions such as ${({\bf a},\,{\bf b})}$ and ${({\bf a},\,{\bf b'})}$ simultaneously. Alice, for example, can make measurements along ${\bf a}$ or ${\bf a'}$, but not along ${\bf a}$ {\it and} ${\bf a'}$ at the same time.

But this inconvenient fact is rather devastating for Bell's argument, because it means that his replacement (\ref{rep}) is illegitimate. Consider a specific run of the EPRB experiment and the corresponding quantity being averaged in (\ref{rep}):
\begin{equation}
{\mathscr A}_{k}({\bf a})\,{\mathscr B}_{k}({\bf b})\,+\,
{\mathscr A}_{k}({\bf a})\,{\mathscr B}_{k}({\bf b'})\,+\,{\mathscr A}_{k}({\bf a'})\,{\mathscr B}_{k}({\bf b})\,-\,
{\mathscr A}_{k}({\bf a'})\,{\mathscr B}_{k}({\bf b'})\,. \label{riy}
\end{equation}
Here the index ${k=1}$ now represents a specific run of the experiment. But since Alice and Bob have only two particles at their disposal for each run, only one of the four terms of the above sum is physically meaningful. In other words, the above quantity is physically meaningless, because Alice, for example, cannot align her detector along ${\bf a}$ and ${\bf a'}$ at the same time. And likewise, Bob cannot align his detector along ${\bf b}$ and ${\bf b'}$ at the same time. What is more, this will be true for all possible runs of the experiment, or equivalently for all possible pairs of particles. Which implies that all\break of the quantities listed below, as they appear in the average (\ref{5}), are un-observable and hence physically meaningless:
\begin{align}
&{\mathscr A}_{1}({\bf a})\,{\mathscr B}_{1}({\bf b})\,+\,
{\mathscr A}_{1}({\bf a})\,{\mathscr B}_{1}({\bf b'})\,+\,{\mathscr A}_{1}({\bf a'})\,{\mathscr B}_{1}({\bf b})\,-\,
{\mathscr A}_{1}({\bf a'})\,{\mathscr B}_{1}({\bf b'})\,, \nonumber \\
&{\mathscr A}_{2}({\bf a})\,{\mathscr B}_{2}({\bf b})\,+\,
{\mathscr A}_{2}({\bf a})\,{\mathscr B}_{2}({\bf b'})\,+\,{\mathscr A}_{2}({\bf a'})\,{\mathscr B}_{2}({\bf b})\,-\,
{\mathscr A}_{2}({\bf a'})\,{\mathscr B}_{2}({\bf b'})\,,  \nonumber \\
&{\mathscr A}_{3}({\bf a})\,{\mathscr B}_{3}({\bf b})\,+\,
{\mathscr A}_{3}({\bf a})\,{\mathscr B}_{3}({\bf b'})\,+\,{\mathscr A}_{3}({\bf a'})\,{\mathscr B}_{3}({\bf b})\,-\,
{\mathscr A}_{3}({\bf a'})\,{\mathscr B}_{3}({\bf b'})\,,  \nonumber \\
&{\mathscr A}_{4}({\bf a})\,{\mathscr B}_{4}({\bf b})\,+\,
{\mathscr A}_{4}({\bf a})\,{\mathscr B}_{4}({\bf b'})\,+\,{\mathscr A}_{4}({\bf a'})\,{\mathscr B}_{4}({\bf b})\,-\,
{\mathscr A}_{4}({\bf a'})\,{\mathscr B}_{4}({\bf b'})\,,  \nonumber \\
&\;\;\;\;.\nonumber \\
&\;\;\;\;.\nonumber \\
&\;\;\;\;.\nonumber \\
&\!{\mathscr A}_{n}({\bf a})\,{\mathscr B}_{n}({\bf b})\,+\,
{\mathscr A}_{n}({\bf a})\,{\mathscr B}_{n}({\bf b'})\,+\,{\mathscr A}_{n}({\bf a'})\,{\mathscr B}_{n}({\bf b})\,-\,
{\mathscr A}_{n}({\bf a'})\,{\mathscr B}_{n}({\bf b'})\,.\nonumber
\end{align}
But since each of the quantities above is physically meaningless, their average appearing on the RHS of (\ref{rep}), namely
\begin{equation}
\Bigl\langle\,{\mathscr A}_{k}({\bf a})\,{\mathscr B}_{k}({\bf b})\,+\,
{\mathscr A}_{k}({\bf a})\,{\mathscr B}_{k}({\bf b'})\,+\,{\mathscr A}_{k}({\bf a'})\,{\mathscr B}_{k}({\bf b})\,-\,
{\mathscr A}_{k}({\bf a'})\,{\mathscr B}_{k}({\bf b'})\,\Bigr\rangle\,,
\end{equation}
is also physically meaningless. That is to say, no physical experiment can ever be performed -- even in principle -- that can meaningfully allow to measure or evaluate the above average, since none of the above list of quantities could have\break experimentally observable values. Therefore the innocuous looking replacement (\ref{rep}) made by Bell is, in fact, illegal. 

On the other hand, it is crucial to note that each of the four averages appearing on the LHS of replacement (\ref{rep}),
\begin{align}
{\cal E}({\bf a},\,{\bf b})\,&=\lim_{\,n\,\gg\,1}\left[\frac{1}{n}\sum_{k\,=\,1}^{n}\,
{\mathscr A}({\bf a},\,{\lambda}^k)\;{\mathscr B}({\bf b},\,{\lambda}^k)\right]\,\equiv\,\Bigl\langle\,{\mathscr A}_{k}({\bf a})\,{\mathscr B}_{k}({\bf b})\,\Bigr\rangle\,, \\
{\cal E}({\bf a},\,{\bf b'})\,&=\lim_{\,n\,\gg\,1}\left[\frac{1}{n}\sum_{k\,=\,1}^{n}\,
{\mathscr A}({\bf a},\,{\lambda}^k)\;{\mathscr B}({\bf b'},\,{\lambda}^k)\right]\,\equiv\,\Bigl\langle\,{\mathscr A}_{k}({\bf a})\,{\mathscr B}_{k}({\bf b'})\,\Bigr\rangle\,, \\
{\cal E}({\bf a'},\,{\bf b})\,&=\lim_{\,n\,\gg\,1}\left[\frac{1}{n}\sum_{k\,=\,1}^{n}\,
{\mathscr A}({\bf a'},\,{\lambda}^k)\;{\mathscr B}({\bf b},\,{\lambda}^k)\right]\,\equiv\,\Bigl\langle\,{\mathscr A}_{k}({\bf a'})\,{\mathscr B}_{k}({\bf b})\,\Bigr\rangle\,, \\
{\text{and}}\;\;\;{\cal E}({\bf a'},\,{\bf b'})\,&=\lim_{\,n\,\gg\,1}\left[\frac{1}{n}\sum_{k\,=\,1}^{n}\,
{\mathscr A}({\bf a'},\,{\lambda}^k)\;{\mathscr B}({\bf b'},\,{\lambda}^k)\right]\,\equiv\,\Bigl\langle\,{\mathscr A}_{k}({\bf a'})\,{\mathscr B}_{k}({\bf b'})\,\Bigr\rangle\,,
\end{align} 
is a perfectly well defined and observable physical quantity. Therefore the bounds (\ref{3}) on their sum are quite harmless. These bounds of ${\{-4,\,+4\}}$, however, have never been violated in any experiment (indeed, nothing can violate them). 

In summary, Bell and his followers derive the upper bound of 2 on the CHSH string of averages by an illegal move. In the middle of their derivation they unjustifiably replace an observable, and hence physically meaningful quantity,  
\begin{equation}
\Bigl\langle\,{\mathscr A}_{k}({\bf a})\,{\mathscr B}_{k}({\bf b})\,\Bigr\rangle\,+\, \Bigl\langle\,{\mathscr A}_{k}({\bf a})\,{\mathscr B}_{k}({\bf b'})\,\Bigr\rangle\,+\,\Bigl\langle\,{\mathscr A}_{k}({\bf a'})\,{\mathscr B}_{k}({\bf b})\,\Bigr\rangle\,-\, \Bigl\langle\,{\mathscr A}_{k}({\bf a'})\,{\mathscr B}_{k}({\bf b'})\,\Bigr\rangle\,,
\end{equation}
with an experimentally un-observable, and hence physically entirely meaningless quantity 
\begin{equation}
\Bigl\langle\,{\mathscr A}_{k}({\bf a})\,{\mathscr B}_{k}({\bf b})\,+\,
{\mathscr A}_{k}({\bf a})\,{\mathscr B}_{k}({\bf b'})\,+\,{\mathscr A}_{k}({\bf a'})\,{\mathscr B}_{k}({\bf b})\,-\,
{\mathscr A}_{k}({\bf a'})\,{\mathscr B}_{k}({\bf b'})\,\Bigr\rangle\,.
\end{equation}
If they do not make this illegitimate replacement, then the upper bound on the CHSH string of averages is 4, not 2. It is mind-boggling why such a blatant mistake has been overlooked by the physics community for over 50 years \cite{local}.

\vfill\eject

One may suspect that the above conclusion is perhaps an artifact of the discrete version, (\ref{exppeu-2}), of the expectation values ${{\cal E}({\bf a},\,{\bf b})}$. Perhaps it can be ameliorated if we considered the CHSH sum (\ref{B1-11-2}) in the following continuous form:
\begin{equation}
\int_{\Lambda}{\mathscr A}({\bf a},\,\lambda)\,{\mathscr B}({\bf b},\,\lambda)\,d\rho(\lambda)\,+\int_{\Lambda}{\mathscr A}({\bf a},\,\lambda)\,{\mathscr B}({\bf b'},\,\lambda)\,d\rho(\lambda)\,+\int_{\Lambda}{\mathscr A}({\bf a'},\,\lambda)\,{\mathscr B}({\bf b},\,\lambda)\,d\rho(\lambda)\,-\int_{\Lambda}{\mathscr A}({\bf a'},\,\lambda)\,{\mathscr B}({\bf b'},\,\lambda)\,d\rho(\lambda)\,, \label{d15}
\end{equation}
where ${\Lambda}$ is the space of all hidden variables ${\lambda}$ and ${\rho(\lambda)}$ is the probability distribution of ${\lambda}$ \cite{Bell}. Written in this form, it is now easy to see that the above CHSH sum of expectation values is both mathematically and physically identical to
\begin{equation}
\int_{\Lambda}\;\Big[\;{\mathscr A}({\bf a},\,\lambda)\,\big\{\,{\mathscr B}({\bf b},\,\lambda)\,+\,{\mathscr B}({\bf b'},\,\lambda)\,\big\}\,+\,{\mathscr A}({\bf a'},\,\lambda)\,\big\{\,{\mathscr B}({\bf b},\,\lambda)\,-\,{\mathscr B}({\bf b'},\,\lambda)\,\big\}\Big]\;\,d\rho(\lambda)\,. \label{d16}
\end{equation}
But since the above two integral expressions are identical to each other, we can use the second expression without loss of generality to prove that the criterion of reality used by Bell is excessively restrictive compared to that of EPR.

{\underbar{Proof}}: As we saw above, (\ref{d16}) involves an integration over fictitious quantities like ${{\mathscr A}({\bf a},\,\lambda)\left\{{\mathscr B}({\bf b},\,\lambda)\pm{\mathscr B}({\bf b'},\,\lambda)\right\}}$. These quantities are not parts of the space of all possible measurement outcomes such as ${{\mathscr A}({\bf a},\,\lambda)}$, ${{\mathscr A}({\bf a'},\,\lambda)}$, ${{\mathscr B}({\bf b},\,\lambda)}$, ${{\mathscr B}({\bf b'},\,\lambda)}$, {\it etc.}; because that space --- although evidently closed under multiplication --- is {\it not} closed under addition. This is analogous to the fact that the set ${{\cal O}:=\{1, 2, 3, 4, 5, 6\}}$ of all possible outcomes of a die throw is not closed under addition. For example, ${3 + 6}$ is not a part of the set ${\cal O}$. But there is also a much more serious physical problem with Bell's criterion of reality. As noted above, the quantities ${{\mathscr A}({\bf a},\,\lambda)\left\{{\mathscr B}({\bf b},\,\lambda)\pm{\mathscr B}({\bf b'},\,\lambda)\right\}}$ are not physically meaningful in {\it any} possible physical world, classical or quantum. That is because ${{\mathscr B}({\bf b},\,\lambda)}$ and ${{\mathscr B}({\bf b'},\,\lambda)}$ can coexist with ${{\mathscr A}({\bf a},\,\lambda)}$ only counterfactually. If ${{\mathscr B}({\bf b},\,\lambda)}$ coexists with ${{\mathscr A}({\bf a},\,\lambda)}$, then ${{\mathscr B}({\bf b'},\,\lambda)}$ cannot coexist with ${{\mathscr A}({\bf a},\,\lambda)}$, and vice versa. But in the proof of his theorem Bell presumes both ${{\mathscr B}({\bf b},\,\lambda)}$ and ${{\mathscr B}({\bf b'},\,\lambda)}$ to coexist with ${{\mathscr A}({\bf a},\,\lambda)}$ simultaneously. That is analogous to being in New York and Miami at exactly the same time. But no reasonable criterion of reality can justify such an unphysical demand. The EPR criterion of reality most certainly does not demand any such thing. 

In conclusion, since the two integrands of (\ref{d16}) are physically meaningless, the bounds of ${-2}$ and ${+2}$ on (\ref{d15}) are\break also physically meaningless. They are mathematical curiosities, without any relevance for the question of local realism.

\parskip 3pt

\acknowledgments
I am grateful to Fred Diether, Michel Fodje, and Albert Jan Wonnink for their efforts to simulate the above model.


\begin{thebibliography}{}

\bibitem[IJTP(2014)]{IJTP} J. Christian, {\sl Macroscopic Observability of Spinorial Sign Changes under ${2\pi}$ Rotations}, Int. J. Theor. Phys., \break DOI 10.1007/s10773-014-2412-2; See also the last two appendices of arXiv:1211.0784. 

\bibitem[Bell(1964)]{Bell} J. S. Bell, Physics {\bf 1}, 195 (1964); {\sl Speakable and Unspeakable in Quantum Mechanics} (CUP, Cambridge, 1987), page 37.
\bibitem[Peres(1993)]{Peres} A. Peres, {\it Quantum Theory: Concepts and Methods} (Kluwer, Dordrecht, 1993), p 161.

\bibitem[Christian(2008)]{Can} J. Christian, {\sl Can Bell's Prescription for Physical Reality be Considered Complete?}, arXiv:0806.3078.

\bibitem[Christian(2014)]{Christian} J. Christian, {\it Disproof of Bell's Theorem: Illuminating the Illusion of Entanglement}, Second Edition (BrownWalker Press, Boca Raton, Florida, 2014). For the latest results see also \url{http://libertesphilosophica.info/blog/}.

\bibitem[Clifford(2003)]{GA} C. Doran and A. Lasenby,
{\it Geometric Algebra for Physicists} (Cambridge University Press, Cambridge, 2003).

\bibitem[Gill(2014)]{Gillprint} R. D. Gill, {\sl Macroscopic Unobservability of Spinorial Sign Changes}, arXiv:1412.2677.

\bibitem[Gill(2012)]{refute} J. Christian, {\sl Refutation of Richard Gill's Argument Against my Disproof of Bell's Theorem}, arXiv:1203.2529.

\bibitem[Wonnink(2015)]{Wonnink} A-J. Wonnink, \url{http://challengingbell.blogspot.co.uk/2015/03/numerical-validation-of-vanishing-of_30.html}, \& C. F. Diether III, \url{http://challengingbell.blogspot.co.uk/2015/05/further-numerical-validation-of-joy.html}.

\bibitem[GAV(2005)]{GAV} L. Dorst, D. Fontijne, and S. Mann, \url{http://geometricalgebra.org/gaviewer_download.html}.

\bibitem[rpub(2014)]{rpub} J. Christian, \url{http://rpubs.com/jjc/84238} [See also \url{http://rpubs.com/jjc/99993} and \url{http://rpubs.com/jjc/105450}].

\bibitem[disproof(2015)]{disproof} J. Christian, {\sl Disproof of Bell's Theorem} (see especially the second version) arXiv:1103.1879v2 [quant-ph] 15 October 2015. 

\bibitem[CHSH(1969)]{CHSH} J. F. Clauser, M. A. Horne, A. Shimony, and R. A. Holt, Phys. Rev. Lett.
{\bf 23}, 880 (1969).

\bibitem[local(2014)]{local} J. Christian, {\sl Local Causality in a Friedmann-Robertson-Walker Spacetime}, \url{http://arXiv.org/abs/1405.2355} (2014). 

\end{thebibliography}
\end{document}